\documentclass[prl,twocolumn,showpacs]{revtex4}
\usepackage{epsfig}
\usepackage{dcolumn}

\begin{document}
\bibliographystyle{apsrev}

\title{Evidence for Scattering-Dependent Multigap Superconductivity in Ba$_8$Si$_{46}$}

\author{Yves Noat, Tristan Cren, Pierre Toulemonde, Alfonso San Miguel, Fran\c{c}ois Debontridder, Vincent Dubost  and Dimitri Roditchev}
\affiliation{Institut des Nanosciences de Paris, Universit\'{e} Pierre
et Marie Curie-Paris $6$ and CNRS-UMR 7588, 4 place Jussieu, 75252
Paris, France}

\affiliation{Institut N\'{e}el, CNRS et Universit\'{e} Joseph Fourier, 25
avenue des Martyrs, BP 166 F-38042 Grenoble cedex 9, France}

\affiliation{Laboratoire de Physique de la Mati\`{e}re Condens\'{e}e et
Nanostructures CNRS UMR-5586, 43 bld du 11 novembre 1918, 69622
Villeurbanne, Lyon, France}

\date{\today}
\begin{abstract}

We have studied the quasiparticle excitation spectrum of the
superconductor Ba$_8$Si$_{46}$ by local tunneling spectroscopy.
Using high energy resolution achieved in
Superconductor-Superconductor junctions we observed tunneling
conductance spectra of a non-conventional shape revealing two
distinct energy gaps, $\Delta_L=1.3\pm0.1$~meV and
$\Delta_S=0.9\pm0.2$~meV. The analysis of tunneling data evidenced
that $\Delta_L$ is the principal superconducting gap while
$\Delta_S$, smaller and more dispersive, is induced into an
intrinsically non-superconducting band of the material by the
inter-band quasiparticle scattering.

\end{abstract}

\pacs{74.25.Gz, 74.72.Jt, 75.30.Fv, 75.40.-s}

\maketitle

In their microscopic theory of superconductivity \cite{BCS} Bardeen,
Cooper and Schrieffer (BCS) predicted the existence of a single gap
$\Delta=\textit{cte}$ in the elementary excitation spectrum of a
superconductor (SC). However, already in late 60s, some SCs from the
A15 family did show anomalies in specific heat that could be
attributed to the presence of several energy gaps
\cite{Brock,Nb3Sn}. The existence of two energy gaps has also been
suggested for Nb, Ta and V \cite{Shen}. Recent discovery of the two
distinct SC gaps in MgB$_2$ reported in specific heat \cite{Bouquet}
and Scanning Tunneling Microscopy/Spectroscopy (STM/STS) experiments
\cite{Giubileo,STSMgB2}, with the excitation spectrum deviating from
the BCS, strongly renewed the interest for non-standard SCs
\cite{Borocarbides,pyrochlore,NbS2,Lortz}.

Among the covalent $sp^3$ materials, like silicon carbide or silicon
(or carbon) diamond, the silicon clathrate Ba$_8$Si$_{46}$ appears
to be a good candidate for a non-conventional superconductivity.
This doped clathrate is formed by covalent Si$_{20}$ and Si$_{24}$
cages sharing their faces and filled with intercalated Ba atoms (for
more details, see \cite{SanMiguel}).  It has been shown that the
superconductivity appearing in Ba$_8$Si$_{46}$ at $T_c = 8.1$~K
\cite{Toulemonde,Kawaji,Yamanaka,Connetable,Tanigaki} is mediated by
phonons and is an intrinsic property of the $sp^3$ network formed by
Si atoms \cite{Connetable}. Eight encaged Ba atoms per unit cell
provide the charge carriers \cite{ToulemondeB} resulting in a
complex band structure with several bands crossing the Fermi level.
Thus, one could expect the superconductivity to appear in two or
more bands, and to depend on the inter-band scattering. Up to now
however, the question remained a subject of controversy: K. Tanigaki
et al. concluded on a conventional BCS superconductivity in
Ba$_8$Si$_{46}$ \cite{Tanigaki}, while tunneling spectroscopy data
evidenced for an anisotropic order parameter \cite{Ichimura}, and
recent results of specific heat measurements were reported to be
consistent with two-gap superconductivity \cite{Lortz}.

\begin{figure}
\begin{center}
\includegraphics[width=7.5cm]{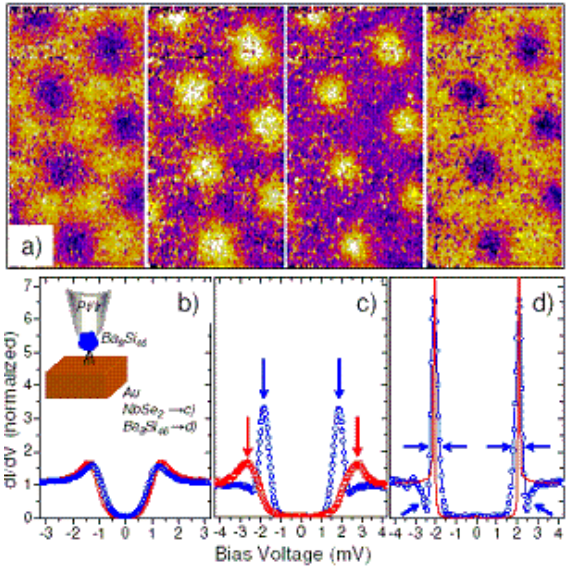}
\end{center}\caption{(Color online)
a) Vortex lattice in $NbSe_2$ at $B=0.165~T$ revealed with
Ba$_8$Si$_{46}$ tips at the biases (from left to right) $-2.0$~mV,
$-0.9$~mV, $+1.0$~mV, $+1.9$~mV shows the  contrast inversion
characteristic to SIS junctions \cite{Kohen}. b) Two typical SIN
tunneling conductance spectra obtained with a Ba$_8$Si$_{46}$ tips
on Au film (red and blue dots). Solid lines: best fits using the
McMillan equations (2) with $\Delta_1^0=0.08~(0.05)$~meV,
$\Gamma_{1}=10~(11)$~meV, $\Delta_2^0=1.2~(1.33)$~meV,
$\Gamma_{2}=1.49~(1.31)$~meV for red (blue) curve; inset - schematic
drawing of the experimental geometry. c) Ba$_8$Si$_{46}$ vs $NbSe_2$
SIS spectra ($T=2.2$~K) reveal two distinct energy gaps in
Ba$_8$Si$_{46}$ (showed with down arrows): $\Delta_S\approx0.8$~meV
(blue curve) and $\Delta_L\approx1.4$~meV (red curve). d)
Ba$_8$Si$_{46}$ vs Ba$_8$Si$_{46}$ SIS tunneling conductance
spectrum (blue curve). Red line: SIS fit with BCS clearly fails to
reproduce the observed large quasiparticle peaks and dips (pointed
with arrows).} \label{Fig1}
\end{figure}

In this Letter we report the local tunneling spectroscopy (TS) of
Ba$_8$Si$_{46}$ performed in the STM geometry. In order to enhance
the energy resolution, we studied
Superconductor-Vacuum-Superconductor (SIS) tunneling junctions
formed by \textit{in-situ} cleaved Ba$_8$Si$_{46}$ grains (used as
STM tips) \cite{Giubileo} and a clean surface of $2H-$NbSe$_2$ or,
alternatively, another grain of Ba$_8$Si$_{46}$, used as a "sample".
Statistical analysis of SIS spectra revealed two energy gaps to
exist in Ba$_8$Si$_{46}$, a large one $\Delta_L=1.3\pm0.1$~meV,
giving a ratio $2\Delta_L/kT_c=3.7 \pm0.3$ matching the BCS value,
and a smaller one $\Delta_S=0.9\pm0.2$~meV. This apparent small gap
is interpreted as being due to the inter-band scattering induced
superconductivity in an intrinsically non-superconducting electronic
band of Ba$_8$Si$_{46}$.

Polycrystalline Ba$_8$Si$_{46}$ samples were synthesized at high
pressure and temperature \cite{Toulemonde,ToulemondeB}. The
Ba$_8$Si$_{46}$ tips were fabricated by gluing a small grain of the
material at the apex of a standard Pt/Ir tip. The grains were then
fractured prior TS experiment in order to expose a clean surface
facing the STM junction (see inset in Fig.1b). In this work, both
ex-situ and in-situ fractured Ba$_8$Si$_{46}$-tips were studied. As
a control of the tip quality, the ability of the tips to image Au or
NbSe$_2$ surfaces was systematically checked. We also visualized the
vortex lattice in NbSe$_2$ and observed the voltage dependent
contrast (Fig.1a) expected for high-quality SIS junctions
\cite{Kohen}. These evidence both the \textit{vacuum tunneling}
regime and the \textit{clean surfaces} of the tunneling electrodes.
The raw $I(V)$ data acquired at tunneling resistances
10-100M$\Omega$ were numerically derived, and the resulting
$dI(V)/dV$ spectra are presented normalized to unity for clarity.

Fig.1b presents typical SIN tunneling conductance spectra obtained
with Ba$_8$Si$_{46}$ tips on Au. The spectra exhibit a smooth
apparent gap $\Delta_{Ba_8Si_{46}}\simeq1.0$~meV with no additional
spectroscopic features visible. The position of the peaks slightly
varies (typically $\pm0.2$~mV) from one grain to another. Such
thermally broadened SIN spectra are quite identical to those
previously reported \cite{Ichimura}, and which were interpreted as
resulting from the anisotropy of the SC gap. A typical
Ba$_8$Si$_{46}$~ -~NbSe$_2$ SIS spectrum is presented in Fig.1c
(doted blue curve). In SIS configuration the peaks appear at
$\Delta_{peak}=\Delta_{NbSe_2}+\Delta_{Ba_8Si_{46}}$. Considering
$\Delta_{NbSe_2}(T=2.2~K)\simeq1.1$~meV leads to
$\Delta_{Ba_8Si_{46}}\simeq0.8$~meV. A typical
Ba$_8$Si$_{46}$~-~Ba$_8$Si$_{46}$ SIS spectrum ($T=2.2$~K) is shown
in Fig.1d (doted blue curve): Its sharpness clearly demonstrates the
effect of enhanced energy resolution in SIS tunneling spectroscopy
with respect to SIN geometry (Fig1b). The best BCS fit (red solid
line in Fig.1d) gives $\Delta_{Ba_8Si_{46}}=1.04$~meV. Thus, both
SIN and SIS data in Fig.1 reveal
$\Delta_{Ba_8Si_{46}}=0.9\pm0.2$~meV, too low to account alone for
the superconductivity in bulk Ba$_8$Si$_{46}$
($2\Delta_{Ba_8Si_{46}}/kT_c=2.6~<~3.52$). Furthermore, the standard
BCS fit (red line in Fig.1d) fails to account for strongly
\textit{broadened quasiparticle peaks} and local minima -
\textit{dips} \cite{footnote1}. Both the low gap energy and the non
conventional shape of the tunneling spectra suggest the existence of
another (hidden) leading SC gap $\Delta_L$ responsible for the
superconductivity in Ba$_8$Si$_{46}$.

The necessity of a precise description of realistic SCs stimulated
an important piece of theoretical effort. In 1959 Suhl, Matthias and
Walker \cite{Suhl} extended the one-band isotropic BCS model to the
case of two energy bands. Such a two-band SC exhibits two distinct
gaps in its excitation spectrum, $\Delta_1$ and $\Delta_2$.
Analytically, the tunneling DOS is the weighted sum of two BCS
spectra $N_{BCS\textit{i}}(E)$ \cite{BCS} with two different
parameters $\Delta_1$ and $\Delta_2$:
\begin{equation}
N_S(E)=\sum_{\textit{i}=1,2}W_\textit{i}~N_{BCS\textit{i}}(E)=\sum_{\textit{i}=1,2}W_\textit{i}~N_\textit{i}(E_F)\frac{|E|}{\sqrt{E^{2}-\Delta_\textit{i}^{2}}}
\end{equation}%
where $W_\textit{i}$ is the tunneling probability and
$N_\textit{i}(E_F)$ - the normal state DOS at the Fermi level in
each band, $\textit{i}=1,2$. Such $N_S(E)$ does not generate any dip
features in SIS spectra and thus, Suhl's model fails to describe our
experimental data.

The inter-band quasiparticle scattering, non accounted for in the
model above, leads to an additional modification of the
quasiparticle excitation spectrum \cite{Schopohl}. The latter may be
calculated considering two initial BCS gaps in each band
$\Delta_1^{0}$ and $\Delta_2^{0}$, and the coupling energies
$\Gamma_{1}$ and $\Gamma_{2}$ accounting for the quasiparticle
scattering from one band to another. Similarly to the case of the
proximity effect in the real space developed by McMillan
\cite{McMillan}, one obtains two coupled equations for the energy
dependent effective gaps $\Delta_1(E)$ and $\Delta_2(E)$:

\begin{eqnarray}
\Delta_1(E)&=&\frac{\Delta_1^{0}+\Gamma_{1}\Delta_2(E)/\sqrt{\Delta_2^2(E)-(E-i\Gamma_{2})^2}}{1+\Gamma_{1}/\sqrt(\Delta_2^2(E)-(E-i\Gamma_{2})^2)} \\
\nonumber\Delta_2(E)&=&\frac{\Delta_2^{0}+\Gamma_{2}\Delta_1(E)/\sqrt{\Delta_1^2(E)-(E-i\Gamma_{1})^2}}{1+\Gamma_{2}/\sqrt(\Delta_1^2(E)-(E-i\Gamma_{1})^2)}
\label{McMillan}
\end{eqnarray}

These equations allow one to calculate the tunneling spectrum of a
two-band SC, by replacing  the constant gaps $\Delta_\textit{i}$ in
(1) by the energy dependent gaps $\Delta_\textit{i}(E)$
\cite{footnote2}.

\begin{figure}
\begin{center}
\includegraphics[width=8cm]{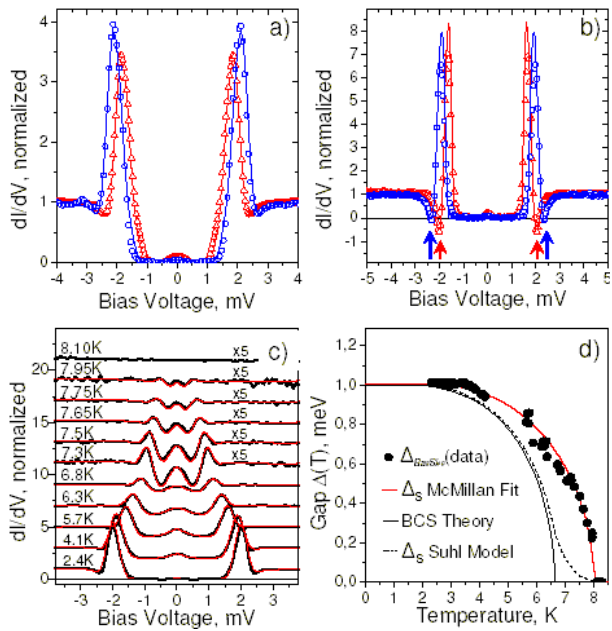}
\end{center}\caption{(Color online)
a) and b) Typical SIS tunneling conductance spectra (red and blue
dots) observed at $T=2.2$~K: a) Ba$_8$Si$_{46}$ vs NbSe$_2$; b)
Ba$_8$Si$_{46}$ vs Ba$_8$Si$_{46}$. In some junctions a negative
tunneling conductance is measured at the dip positions (arrows).
Solid lines: best SIS fits with McMillan equations (2) obtained with
the following parameters: Red (blue) curve in a):
$\Delta_1^0=0.0~(0.07)$~meV, $\Gamma_{1}=2.2~(3.54)$~meV,
$\Delta_2^0=1.1~(1.34)$~meV, $\Gamma_{2}=0.17~(0.26)$~meV, the exact
DOS of NbSe$_2$ was taken from \cite{Rodrigo}; Red (blue) curve in
b): $\Delta_1^0=0.0~(0.0)$~meV, $\Gamma_{1}=2.6~(3.7)$~meV,
$\Delta_2^0=1.1~(1.23)$~meV, $\Gamma_{2}=0.18~(0.40)$~meV; c)
Temperature dependence of Ba$_8$Si$_{46}$ vs Ba$_8$Si$_{46}$ spectra
(thick black lines) and their fits with (2) (thin red lines). The
spectra are shifted for clarity; d) Comparison of the SIS data in c)
with different models.} \label{Fig2}
\end{figure}

In Fig.2a,b we present SIS tunneling spectra, Fig.2c showing their
evolution with temperature. The enhanced energy resolution of SIS
spectroscopy revealed the apparent gap $\Delta_{Ba_8Si_{46}}$ to
vanish exactly at the $T_c$ of the bulk material, thus evidencing
its relation to the bulk SC (Fig.2d). The SIS fits using eqs. (2)
(solid lines in Fig.2) reproduce in fine details the shape of the
SIS tunneling conductance data, with different counter electrode
materials (in the case of ${NbSe_2}$ , Fig.2a, the exact DOS was
taken from \cite{Rodrigo}) and at various temperatures (red lines in
Fig.2c). A perfect agreement is also achieved when fitting SIN
spectra (solid lines in Fig.1b). By analyzing the fitting parameters
of different tunneling spectra we noticed the following common
features: i) for all spectra initial BCS gap in the band
$\textit{i}=1$ is $\Delta_1^{0}\approx 0$ while
$\Delta_2^{0}=1.3\pm0.1$~meV. ii) the interband scattering
parameters, $\Gamma_{1}$ and $\Gamma_{2}$ strongly vary from one
junction to another, but their ratio remains almost fixed
$\Gamma_{1}/\Gamma_{2} \sim$10 (see also Fig.3d). iii) for most of
the studied junctions the tunneling contribution of the band
$\textit{i}=1$ was $W_{1}=1$ and that of $\textit{i}=2$ was
$W_{2}=0$, i.e. we needed only the term $N_{S1}(E)$ to fit the
tunneling spectra. Thus, only two really free parameters remained,
$\Delta_2^{0}$ and $\Gamma_{1}(\Gamma_{2})$. Moreover, the fits in
Fig.2c were generated considering only one free parameter,
$\Delta_2^0$, the second being evaluated at $T=2.2$~K
($\Gamma_{1}=10~\Gamma_{2}=6$~meV) and kept fixed.

Let us now discuss the physics behind such a surprisingly nice
agreement between the data and fits, and analyze the consequences of
the above mentioned common features. The condition
$\Delta_1^{0}\approx 0$~meV implies that the superconductivity in
the band $\textit{i}=1$ is fully induced: a small gap opens in
$N_{S1}(E)$ due to the coupling with the leading superconducting
band $\textit{i}=2$ where the pairing amplitude
$\Delta_2^{0}\simeq1.3$~meV is found close to the value 1.4~meV
estimated from specific heat measurements \cite{Lortz}. The observed
fluctuations of the induced small gap $\Delta_S$ from one junction
to another, $\pm0.2$~meV, are attributed to the variations of both
the scattering rates $\Gamma_{\textit{i}}$ and the pairing amplitude
of the initial gap $\Delta_2^0$. Besides, a larger scattering due to
the surface disorder may explain the differences between $\Delta_S$
observed in TS data and the one ($0.35$~meV) found in specific heat
measurements \cite{Lortz}. Indeed, the value $\Delta_S=0.35$~meV can
be easily calculated with (2) considering a smaller
$\Gamma_{1}=0.6$~meV in the bulk.

A clear linear correlation between $\Gamma_{1}$ and $\Gamma_{2}$ is
evidenced in Fig.3d, with $\Gamma_{1}/\Gamma_{2} \sim$10. This
result is consistent with the scattering in a two-band material: The
interband scattering events $1\rightarrow2$ and $2\rightarrow1$
should be equal and therefore the condition
$\Gamma_{1}/\Gamma_{2}=N_2(E_F)/N_1(E_F)$ must be fulfilled
\cite{Schopohl}. The ratio $\sim$10 we find is in very good
agreement with the value $N_2(E_F)/N_1(E_F)=9$ determined from
specific heat measurements \cite{Lortz}, that strongly supports our
model. Concerning the variations of $\Gamma_{1}$ and $\Gamma_{2}$
observed in different experiments, we relate them to the specific
scattering conditions due to the local surface disorder, realized
for each junction \cite{footnote3}.

In Figs.3a and 3b we show the partial DOSs $N_{S\textit{i}}(E)$
calculated for bands $\textit{i}=1$ and $\textit{i}=2$ respectively.
We took $\Delta_1^{0}=0$~meV and $\Delta_2^{0}=1.3$~meV, different
curves corresponding to the coupling parameters $\Gamma_{1}$ varying
from 0.5 to 15~meV with a constant ratio $\Gamma_{1}/\Gamma_{2}=10$.
As a result of the inter-band quasiparticle scattering, a gap
$\Delta_S$ is induced in the first band where it does not exist at
zero coupling. The DOS shape in the first band is very peculiar
(Fig.3a): The quasiparticle peaks are broadened and there are some
shoulders at the position of the large gap peaks. This peculiar
shape is responsible of the dips in SIS spectra, experimentally
observed. Broad peaks and dips are also observed in high-$T_c$ SCs
and attributed to the coupling with a collective excitation mode
\cite{Zasadzinski}. Their amplitudes and the characteristic energies
(normalized to the $T_c$) are however much higher as compared to our
case. In the second band, Fig.3b, the initially pure BCS DOS evolves
with the coupling: A mini-gap (often called 'excitation gap')
appears at the same position that the induced gap in the first band.
The evolution of the gaps $\Delta_L$ and $\Delta_S$ in the partial
DOSs with the coupling parameter $\Gamma_{1}$ is presented in
Fig.3c. Both gaps vary strongly at moderate $\Gamma_{1}$ and reach
the same asymptotic value for very large coupling, where one
recovers a simple one-gap BCS DOS.

\begin{figure}
\begin{center}
\includegraphics[width=7.cm]{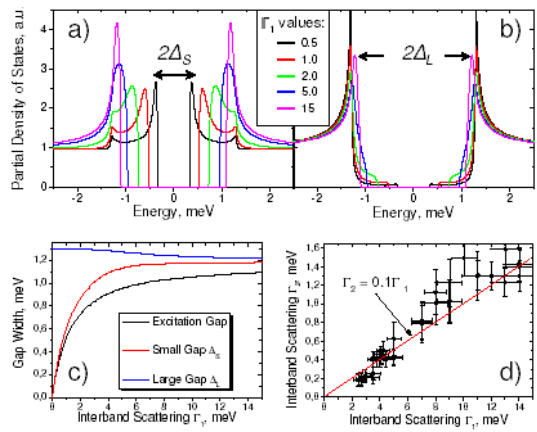}
\end{center}\caption{(Color online)
Partial DOSs in the band 1 in (a) and 2 in (b) calculated within the
McMillan model with $\Delta_1^{0}=0$, $\Delta_2^{0}=1.3$~meV,
$\Gamma_1=0.5, 1, 2, 5, 15 $~meV, $\Gamma_2=\Gamma_1/10$. (c)
Evolution of the excitation gap and apparent gaps $\Delta_L$,
$\Delta_S$ with the scattering rate $\Gamma_1$, corresponding to (a)
and (b). (d) Scatter plot of the fit parameters $\Gamma_2$ vs
$\Gamma_1$. Red line: linear dependence $\Gamma_2=\Gamma_1/10$.}
\label{Fig1}
\end{figure}

In the search for a direct evidence of the large gap $\Delta_L$, we
studied several tens of Ba$_8$Si$_{46}$ tips and acquired the SIS
spectra in thousands of locations, similar to what was done in the
case of $MgB_2$ \cite{Giubileo_EPL}. We expected that the surface
defects such as large steps, protrusions, holes would provide
various tunneling conditions. In some measurements we indeed
observed very different still reproducible SIS spectra (red curve in
Fig.1c), while keeping the tunneling conditions unchanged. They show
no dips, as expected for SIS spectra revealing the large gap,
Fig.3b. Furthermore, the gap energy $\Delta_L\approx1.4$~meV
corresponding to these spectra is in a good agreement with the large
gap deduced from specific heat measurements by Lortz et al.
\cite{Lortz}.

It is not fully established for the moment why the spectra revealing
directly the large gap are so rare. In the case of $MgB_2$ a similar
statistics was observed \cite{Giubileo_EPL} and attributed to the
low probability of tunneling into the two-dimensional $\pi$-band. It
is not clear if such an assumption holds for Ba$_8$Si$_{46}$. We
know however that in Ba$_8$Si$_{46}$ most of the DOS belongs to
Ba-Si hybridized states located inside the Si cages
\cite{Tanigaki,ToulemondeB}. It couples well to Ba phonon rattling
modes that contribute to the SC pairing \cite{Lortz,Tse}. Hence, we
may speculate that the electronic states responsible for the
superconductivity are somehow confined inside the clathrate cages,
and the amplitude of the corresponding evanescent waves outside the
material is thus weak, leading to a tiny contribution into the
tunneling. Consequently, most of the tunneling current comes from
the outer-cage Si orbitals where the superconductivity is purely
induced.

Finally, we have studied the quasiparticle excitation spectrum of
the superconductor Ba$_8$Si$_{46}$ by local tunneling spectroscopy.
Owing high energy resolution achieved in SIS junctions we resolved
fine spectroscopic features that cannot be accounted for by BCS. The
SIS spectra evidence the existence of two distinct energy gaps,
$\Delta_L=1.3\pm 0.1$~meV and $\Delta_S= 0.9\pm0.2$~meV, that are
interpreted in terms of a two-band superconductivity in
Ba$_8$Si$_{46}$ characterized by a leading gap
$2\Delta_L/kT_c=3.7\pm0.3$ and another, smaller coupling-dependent
gap $\Delta_S$, reflecting the superconductivity induced in an
intrinsically non-superconducting electronic band.


\begin{thebibliography}{00}
\bibitem{footnote1}
Note that Ba$_8$Si$_{46}$ - Ba$_8$Si$_{46}$ junctions do not contain
any host material (as it is the case of Ba$_8$Si$_{46}$ - NbSe$_2$
junctions). Thus, the observed deviations from BCS behavior are
intrinsic to Ba$_8$Si$_{46}$.
\bibitem{footnote3}
Note that the surface itself is a topological defect with respect to
the bulk material, due to the crystal symmetry breaking. I could in
principle result in a new electronic band, present only at the
surface (as it is the case of Cu(111) or Ag(111), for instance). The
superconductivity could be induced to this band by the coupling with
the bulk SC, and thus, $\Delta_S$ could, in principle, represent the
surface state SC gap which would not extend to the bulk. Such
hypothesis is generally valid for the surface of any SC, and cannot
be ruled out even for in-situ prepared junctions. However, we did
not find any evidence for such a surface band.
\end{thebibliography}
\end{document}